\documentclass{article}

\usepackage{arxiv}

\usepackage[utf8]{inputenc} % allow utf-8 input
\usepackage[T1]{fontenc}    % use 8-bit T1 fonts
\usepackage{hyperref}       % hyperlinks
\usepackage{url}            % simple URL typesetting
\usepackage{booktabs}       % professional-quality tables
\usepackage{amsfonts}       % blackboard math symbols
\usepackage{nicefrac}       % compact symbols for 1/2, etc.
\usepackage{microtype}      % microtypography		% Can be removed after putting your text content
\usepackage{graphicx}
\usepackage{apacite}
\usepackage{doi}
\usepackage{amsmath}  
\usepackage{subcaption}

\title{Contagion on Financial Networks: An Introduction}

\author{ \href{https://orcid.org/0000-0000-0000-0000}{\includegraphics[scale=0.06]{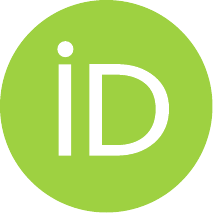}\hspace{1mm}Sunday Akukodi Ugwu}\thanks{ Department of Mathematics, Michael Okpara University of Agriculture Umudike, Nigeria\\ Mfano Africa-Oxford Mathematics 2023 sponsored mini-project\\ Contact: sunday.ugwu@mouau.edu.ng}
}

\renewcommand{\shorttitle}{Contagion on Financial Networks: An Introduction}

\hypersetup{
pdftitle={Contagion on Financial Networks: An Introduction},
pdfsubject={q-fin.MF},
pdfauthor={S. A. Ugwu},
pdfkeywords={Keywords: financial contagion; banks' solvency; models of network; Erdős–Rényi model},
}

\begin{document}
\maketitle

\begin{abstract}
This mini-project models propagation of shocks, in time point, through links in connected banks. In particular, financial network of 100 banks out of which 15 are shocked to default (that is, 85.00\% of the banks are solvent) is modelled using Erdős and Rényi network -- directed, weighted and randomly generated network. Shocking some banks in a financial network implies removing their assets and redistributing their liabilities to other connected ones in the network. The banks are nodes and two ranges of probability values determine tendency of having a link between a pair of banks. Our major finding shows that the ranges of probability values and banks' percentage solvency have positive correlation.	

\medskip
\textbf{keywords:} Financial contagion . banks' solvency . models of network . Erdős–Rényi model
\end{abstract}

% keywords can be removed
   %\keywords{financial contagion \and  banks' solvency \and models of network \and Erdős–Rényi model}

\section{Introduction}
\noindent Financial mathematics which explores stochastic differential equation (SDE), has overlaps with mathematics, statistics, economics and finance. The applied mathematics has found its another overlap with network science. ``Contagion on Financial Networks: An Introduction", the title of this mini-project, is an overlap between network science and financial mathematics.

\medskip
\noindent In financial networks, quantities of each bank's assets (claims) and liabilities (obligations), which can be observed in its exposures with other banks, determine its solvency or default rate when financial crises occur. A solvent bank is one whose assets exceeds its liabilities. A random-directed-weighted network can be used to model such financial network. Each bank's assets are modelled as incoming-weighted links and outgoing-weighted links model its liabilities.

\medskip
\noindent This mini-project models financial contagion -- propagation of shocks (in time point) from any bank (a node) to another in a financial network. As contagious diseases (for example, active tuberculosis (TB)) can spread from some infected persons to others, so financial distress can propagate from some banks to others in a financial system. Some banks are shocked to default -- removing their assets and redistributing their liabilities to others in the network. The mini-project explores how probability of connectivity (tendency of having a link between a pair of banks) in financial network can affect any bank's solvency or default rate. 

\medskip
\noindent One of the earliest contributors to contagion in financial structure emanating from direct linkages is \cite{allen2000financial}. Their findings reveal that complete network structure has capacity of withstanding economic crises because impact of such shocks to the system can be evenly redistributed among other banks. No financial contagion results afterwards.

\medskip
\noindent A paper from \cite{gai2010contagion} is the primary source of this mini-project. The authors develop an analytical model of contagion in ﬁnancial networks with arbitrary structure. They apply random networks which have arbitrary degree distributions, while assuming a uniform (Poisson) random network in which each directed link is generated with independent probability $p$. They explore how aggregate and idiosyncratic shocks -- wiping
out any bank's external assets, can led to contagion in financial networks. The result of their analysis is that although contagion tendency may be low, it can spread widely which shows that financial system has a \textit{robust-yet-fragile} tendency.

\subsection{Aim and Scope of the Mini-Project} 

\noindent We aim at modelling condition for each bank's solvency and investigate the effects of probability of connectedness of any bank when it is shocked to default. 

\medskip
\noindent Condition for any bank to be solvent is \cite{gai2010contagion}

\begin{equation}
(1-\phi)\textbf{A}_{i}^{IB}+q\textbf{A}_{i}^{M}-\textbf{L}_{i}^{IB}-\textbf{D}_{i} > 0,
\label{fig: solvencyeq}
\end{equation}

where:

\begin{itemize}
    \item $\textbf{A}_{i}^{IB}$ denotes the interbank assets ($\textbf{A}_{i}^{IB}$ = 0 if a bank has no incoming links),
    \item $\textbf{A}_{i}^{M}$ denotes the illiquid external assets such as each bank's mortgages,
    \item $\textbf{L}_{i}^{IB}$ denotes the interbank liabilities which are endogenously determined,
    \item $\textit{i}$ denotes specific bank being considered,
    \item interbank exposures of bank $\textit{i}$ deﬁne the links with other banks,
    \item $\phi$ is the fraction of banks with obligations to any bank $\textit{i}$ that has defaulted,
    \item $q$ is the resale price of the illiquid asset, $q \leq 1$ in the event of asset sales by any bank in default, but $q = 1$ if there are no ‘ﬁre
sales’ and
    \item $\textbf{D}_{i}$ denotes the customers' deposits which are exogeneously determined.
\end{itemize}

\noindent The model we implement is 

\begin{equation}
 \textbf{A}_{i}^{IB} + \textbf{A}_{i}^{M} = \textbf{AS}_{i}^{T} \ \textnormal{and} \
\textbf{L}_{i}^{IB} + \textbf{D}_{i} = \textbf{LI}_{i}^{T}.
\label{fig: solvencyImp}
\end{equation}

\begin{itemize}
    \item $\textbf{AS}_{i}^{T}$ is each bank's total assets,
    \item $\textbf{LI}_{i}^{T}$ is its total liabilities and
    \item $\textbf{A}_{i,}^{IB}\ \textbf{A}_{i,}^{M} \ \textbf{L}_{i}^{IB}$ and $\textbf{D}_{i}$ are as defined in equation \eqref{fig: solvencyeq}; $\textbf{A}_{i}^{IB}$ and $\textbf{L}_{i}^{IB}$ are from banks while $\textbf{A}_{i}^{M}$ and $\textbf{D}_{i}$ are randomly constructed.
\end{itemize}

\noindent A solvent bank is one whose net assets, the bank’s capital buffer $K_{i} > 0;$ that is, 

\begin{equation}
    \textbf{AS}_{i}^{T}-\textbf{LI}_{i}^{T} = K_{i} > 0.
\end{equation}

\subsection{Outline}
\noindent Apart from the introduction, this mini-project has 4 other sections. 
\begin{itemize}
    \item \textbf{Section 2:} In this section, an overview of network science as practical applications of graph theory is done. Networks are synonymous with graphs. More attention is paid to Erdős and Rényi network, a random and directed network, which is implemented in this mini-project. 
    \item \textbf{Section 3:} The section provides idea of propagating or spreading processes across nodes through links in networks. Inferences are drawn from random networks, cascade effects in networks and financial contagion. 
    \item \textbf{Section 4:} This section contains the implementation details of the Erdős–Rényi model (E-R model) and their results. The implementation explores percentage solvency and default which depend on the probability values for having link between a pair of banks. Interconnectedness of the financial network determines how far shocks propagate.
    \item \textbf{Section 5:} Here, the summary of the E–R model is done, its importance and possible future work of applying stochastic methods in examining random network are stated.  
\end{itemize}

\section{Introduction to Network Science}

\noindent This section presents an overview of network science as practical applications of graph theory. Networks are graphs. Attention is given to random networks whose properties are compared with real-world networks. In particular, Erdős and Rényi network, a random and directed network, the model of network implemented in this mini-project, is described.

\vskip 0.2in
\begin{definition}[A Network]
 A Network is conceptualized in mathematics as a graph $G := G(V,E)$. $V$ is a set of nodes (vertices or points) which is connected by $E$, set of edges (links, lines) \cite{borgatti2018analyzing}.   
\end{definition}  
\noindent Nodes are drawn as dots or circles and can represent objects, places, individuals, institutions or ideas. 
\begin{figure}[H] % or [h] for less strict placement
\includegraphics[width=0.50\textwidth]{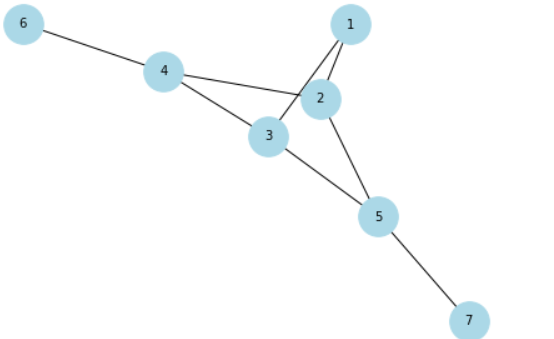} \centering
\caption{An Undirected Network with 7 Nodes and 8 Links}
\label{fig:networkUnd}
\end{figure}

\noindent A directed network is a graph whose links are represented by arrows (cf. Figure \eqref{fig:erdosrenyi}) and when assigned weights, can denote cost, assets, liabilities or distance. An undirected network (cf. Figure \eqref{fig:networkUnd}) has two-directional connections between each pair of nodes and can denote a two-way relationship, say link from node $i$ to node $j$ and vice versa. The maximum number of links of an undirected network on $N$ number of nodes is 
\begin{equation}
    \binom{N}{2} = \frac{N(N-1)}{2} \ \textnormal{edges}
    \label{fig: maxedges}
\end{equation}

\noindent  and twice as many, in a directed network \cite{editionreinhard}. 

\subsection{Graph Theory and Networks}

\noindent Given that $V(G)$ and $E(G)$ denote sets of vertices and edges of graph $G$ respectively. Let nodes $i,j\in V(G)$. If $(i,j)\in E(G),$ then $i,j$ are said to be adjacent or neighbours and $a_{i,j}$ denotes adjacency or connectivity between the 2 nodes, $i,j$ \cite{editionreinhard}. Adjacency implies that there is a relationship between the pair of nodes.

\begin{figure}[H] % or [h] for less strict placement
\includegraphics[width=0.50\textwidth]{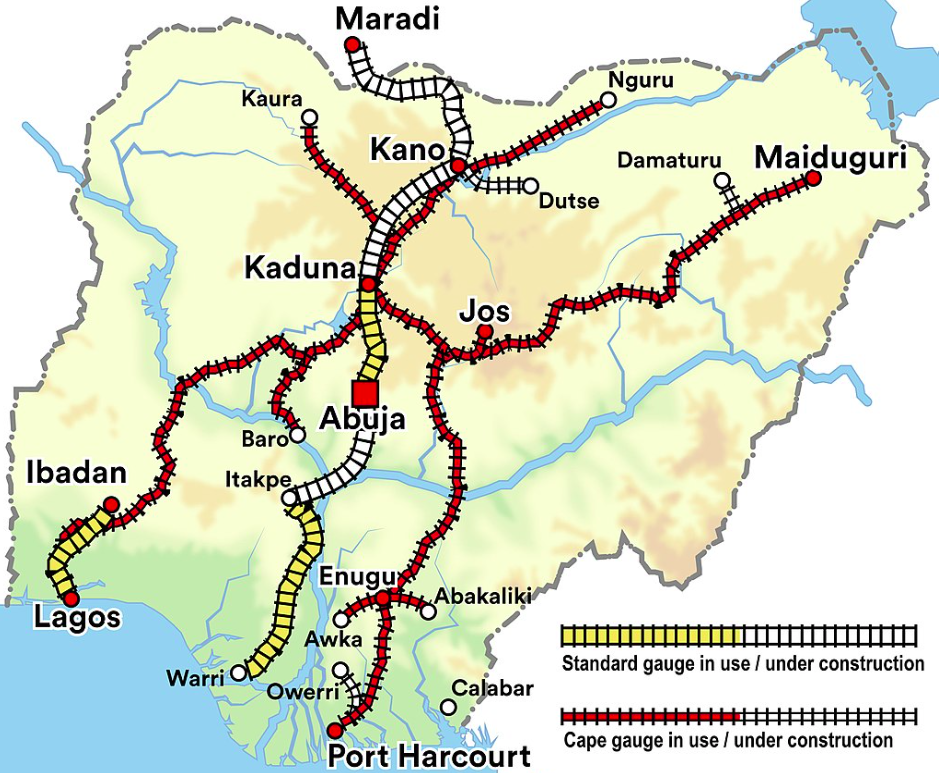} \centering
\caption [Railway]{Railway Lines and Towns in Nigeria form a Network}
\label{fig:railnigeria}
\end{figure}

\noindent A network can exist when there is, at least, a link between a pair of nodes. Figure \eqref{fig:railnigeria} \cite{wikipedia} shows railway network in Nigeria which consists of towns that are linked by rail lines. 

\medskip
\noindent A network is directed when $a_{i,j} \neq a_{j,i}$ but undirected if $a_{i,j} = a_{j,i}$; it is unweighted adjacency if $a_{i,j}$ takes only 1 or 0 but when it takes real values between 0 and 1, it is a weighted adjacency \cite{hong2014recent}. Adjacency matrix is matrix that shows how the network is connected. The matrix enables us to deduce quickly whether 2 nodes share a relationship or not. It is symmetric for an undirected network. We examine random networks and their types.

\subsection{Random Networks}

\noindent A random network is a graph where $N$ labelled nodes are connected with probability $p$. It is an $N$ labelled nodes which are connected with links $L$ that are randomly placed \cite{erdHos1960evolution}\cite{erdHos1961strength}.

\medskip
\noindent Our study of random network is to provide a standard for examining the real-world network -- understand the properties of real-world network.

\subsubsection{Erdős-Rényi Model}

\begin{figure}[H] % or [h] for less strict placement
\includegraphics[width=0.50\textwidth]{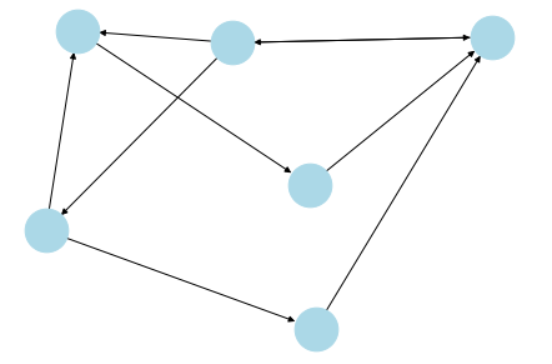} \centering
\caption [Graph]{Directed Erdős and Rényi Network with 6 Nodes and $p =0.20$}
\label{fig:erdosrenyi}
\end{figure}

\noindent Erdős and Rényi random network is the simplest and one of the earliest random network models \cite{posfai2016network}. Figure \eqref{fig:erdosrenyi} is an Erdős and Rényi network with directed, independently and randomly assigned links.

\noindent $G(N,L)$ model and  $G(N,p)$ model (where $N$, $L$ and $p$ represent number of nodes, number of links and probability respectively), are 2 ways of generating E–R model \cite{nobari2011fast}. In the case of $G(N,L)$ model, a network is chosen uniformly at random from sets of $N$ nodes and $L$ links. The nodes are labelled and networks are generated by permuting distinct nodes. For instance, $G(3,2)$ network is a random graph with 3 labelled nodes and 2 links. 3 possible networks can be generated with each of them having probability $\frac{1}{3}$ of occurring in each case. For $G(N,p)$ model, a network is generated by linking nodes randomly with independent probability $p$. If $p$ is high then network of high density of average degree is generated. We adopt directed version of $G(N,p)$ model in this mini-project.

\medskip
\noindent Finding the probability of having number of links $L$ (which is denoted by $\mathbb{P}(L)$), in a network of $N$ number of nodes and probability $p$ is based on probability mass function (PMF) of binomial distribution.
\begin{equation}
  \mathbb{P}(L) = \binom{{N\choose 2}}{L}p^{L}(1-p)^{\frac{N(N-1)}{2}-L},
  \label{fig: pofL}
\end{equation}
\noindent where:\par 
\begin{itemize}
    \item ${N\choose 2}$ is the maximum number of links in a network (cf.equation \eqref{fig: maxedges}),
    \item $p^{L}$ is the probability of having $L$ links,
    \item $(1-p)^{\frac{N(N-1)}{2}-L}$ is the probability of missing $\binom{N}{2}-1$ links and
    \item $\binom{{N\choose 2}}{L}$ is the varying ways that $L$ can be chosen  among all possible links in the network \cite{posfai2016network}.
\end{itemize}

\noindent The expected number of links in E-R model is ${N\choose 2}p.$ Since each node is independently linked with any other node with the probability $p$, its degree distribution is a binomial distribution
\begin{equation}
    \mathbb{P}(z) = \binom{N-1}{z} = p^{z}{(1-p)}^{N-1-z},
    \label{fig: binoDis}
\end{equation}
\noindent where $z$ is the degree of each node and the probability $p$ is to be sufficiently small so that $z_{\textnormal{av}}$, the average degree of the nodes converges to a positive constant in equation \eqref{fig: binoDis} \cite{curien2022erdos}. Poisson distribution \begin{equation}
    \mathbb{P}(z) = \frac{(z_{\textnormal{av}})^{z}\cdot e^{-z_{\textnormal{av}}}}{z!}
    \label{poissDis}
\end{equation}
\noindent is its approximation.

\noindent
\noindent Since general aim of random network models is to explain and model the properties of real-world networks through probabilistic or stochastic methods \cite{chen2022properties}, let us consider a stochastic method.

\subsubsection{Erdős-Rényi Mixture Model}

\noindent An extension of E-R network model is stochastic block model (SBM) -- Erdős-Rényi mixture model, which provides a framework for modelling networks in communities \cite{wang2017likelihood}. The SBM is used to examine clustering in networks and their latent structure \cite{herlau2014infinite}. It depends on 3 parameters: $N$ number of nodes, probability vector $p$ of dimension $K$ blocks' partitions  and $W$ connectivity matrix (a symmetric matrix with entries in [0,1]) -- connectivity among communities. Probability of intra-connectivity is higher than inter-connectivity probability.  

\medskip
\noindent In SBM$(N,p,W)$, there are partitioning of nodes in blocks of arbitrary sizes and any 2 nodes are linked independently, with a probability $p$ which depends on the blocks' partitioning  \cite{holland1983stochastic}. The blocks are used to distinguish the nodes and stochastic generalization of the blocks is provided by the model \cite{holland1983stochastic}. 

\subsubsection{Configuration and Preferential Attachment Models}
\noindent Configuration model and preferential attachment model are random networks models \cite{chen2022properties}. Configuration model is a generalization of the Erdős-Rényi random graph to the case of having parallel links (multiple links), self loop (a link that connects a node to itself) and by randomly assigning links to match given degree sequence \cite{newman2003structure}. In the preferential attachment model, more weights are placed on nodes with high degree as nodes are connected randomly at a time, to existing nodes. Let us discuss properties of random networks.

\subsection{Properties of Random Networks}

\noindent Real-world and random networks often behave differently. We briefly discuss 3 properties of random networks, especially properties of E-R network, which include degree distribution, clustering and average path length.

\subsubsection{Degree Distribution}
\noindent For E-R network, the average degree of each node follows a Poisson distribution but average degree in real network is not Poisson \cite{posfai2016network}. Why? E-R network is generated by choosing a large number of nodes $N$ with small probability value of having a link between a pair of nodes. So, the degree of any node is drawn from a binomial distribution with large $N$ draws with very small chance for success which is approximated by the Poisson distribution \cite{curien2022erdos}.

\subsubsection{Clustering Coefficient}
\noindent  Clustering coefficient is used to measure the connectedness of neighbours of adjacent nodes. Due to some variants that are encountered in describing clustering, let us focus on local clustering coefficient of a node -- the portion of the neighbours of the node which are also neighbours (adjacent to) themselves \cite{breschi2005clusters}.
\begin{definition}[Local Clustering Coefficient]
Let G be a graph (random or non-random). The number of triangles through a node v is given by
\begin{equation*}
    \Delta_{G}(v) = \sum_{u,w\in V(G)}^{}\mathbbm{1}_{\{u,v\}\in E(G)}\mathbbm{1}_{\{v,w\}\in E(G)}\mathbbm{1}_{\{w,u\}\in E(G)} = 2N_{v}
\end{equation*}
  \noindent and the clustering coefficient of a particular node v is given by \begin{equation*}
      CC_{G(v)} = \frac{\Delta_{G(v)}}{d_{G(v)}(d_{G(v)}-1)}.
  \end{equation*}
  \noindent The \textbf{local clustering coefficient} of G is 
  \begin{equation}
      \overline{CC}_G = \frac{1}{\vert V(G)\vert}\sum_{v\in V(G)}^{}{CC}_G(v) = \frac{1}{\vert V(G)\vert}\sum_{v\in V(G)}^{}\frac{\Delta_{G}}{d_{G(v)}(d_{G(v)}-1)},
  \end{equation}
\end{definition}
\noindent where: \par
\begin{itemize}
    \item $\vert V(G)\vert$ is the number of nodes in the graph G,
    \item $d_G(v)$ is the degree of node $v$ in G,
    \item $u,v$ and $w$ are nodes of G and
    \item $N_{v}$ is the number of links between neighbours of $v$ \cite{chen2022properties}.
\end{itemize}

\medskip
\noindent The \textbf{local clustering coefficient}, $\overline{CC} \in [0,1]$ in real network is often high -- not close to zero, unlike in the random networks \cite{chen2022properties}. High clustering coefficient indicates that there are numerous strong ties in the network while low clustering coefficient indicates otherwise.  Average path length of a random network depends on degree of connectivity in the network. We discuss the average path length.

\subsubsection{Average Path Length}
\noindent A path $P = (V,E)$ in a network $G$ is sets of nodes and links of the form \begin{equation*}
    V = \{v_{0},v_{1},...,v_{k}\} \ \textnormal{and} \ E = \{v_{0}v_{1},v_{1}v_{2},...,v_{k-1}v_{k}\}.
\end{equation*}

\begin{figure}[H] % or [h] for less strict placement
\includegraphics[width=0.50\textwidth]{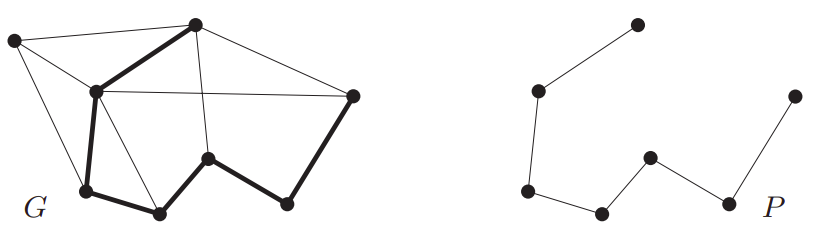} \centering
\caption [Graphp]{A Path P in Network G}
\label{fig:pathinG}
\end{figure}

\begin{itemize}
    \item each node $v_{i}$ is distinct and
    \item nodes $v_{0}$ and $v_{k}$, the end-links or the ends, are linked by $P$ \cite{editionreinhard}.
\end{itemize}
\noindent Length of a path is its number of links. Figure \eqref{fig:pathinG}\cite{editionreinhard} is a path of length 6.  

\noindent The average path length (average distance) of random network refers to mean of all the path lengths in the network. Study of average path length helps us to know how many links one can transverse in a network. Average path length, $L$ for both real-world and random networks is the same and it is given by 
\begin{equation}
    L \approx \frac{\textnormal{log}N}{\textnormal{log} z_{\textnormal{av}}}
    \label{fig: pathLen}
\end{equation}
\noindent given that the probability $p> \mathcal{O}(N\log{}N)$ as ${N \to \infty}$, in order to have connected network \cite{posfai2016network}. $z_{\textnormal{av}}$ is the average degree of nodes in the random network.

 \section{Spreading Processes on Networks}
 \noindent As a background to financial contagion, this section describes the notion of propagation, diffusion or transmission of information or something across nodes through links on networks. Random walks are model of spreading processes on networks. Shocks in financial network propagate in time point due to cascade effects in financial networks.

 \vskip0.2in
\subsection{Basics of Spreading Processes on Networks}
\noindent Spreading phenomena such as spreading of diseases, information or computer viruses etc.\cite{liu2019spreading} that occupy particular nodes on each time point can spread or propagate on particular network since their nodes are connected. Think of a greenish-liquid-pigment which some of its drops are put into a bucket of water, within some intervals of time the water in the bucket is decolourized. When a phenomenon occurs at a node of a network, such can spread from the node to other nodes in the network through the links. 

\medskip
\noindent How much does completeness or incompleteness of networks affect propagation? Are all the nodes have equal capacity of dissemination on the networks? How resistance are some nodes of particular network to propagation? How much of quantity per node of "object of spread" at time $t$ remains at subsequent time $t+1$? We consider equation \eqref{fig: spreadingPr} and some descriptions therein \cite{zhukov2015diffusion}.
\begin{equation}
  \phi_{i}(t+1) = \phi_{i}(t) + \sum_{j}A_{ij}(\phi_{j}(t)-\phi_{i}(t))c\delta t,
  \label{fig: spreadingPr}
\end{equation}
\noindent where:\par
\begin{itemize}
    \item $\phi_{i}(t)$ is the quantity per node at time $t$,
    \item $\phi_{i}(t+1)$ is the quantity remaining of the "object of spread" at subsequent time $t+1$,
    \item $\phi_{j}(t)-\phi_{i}(t)$ is the difference between the value of a node $j$ and its neighbour $i$,
    \item $A_{ij}$ is the adjacency matrix -- in every node, influence from the neighbour is added and
    \item $c\delta t$ refers to time interval ($\delta t$ scaled by $c$) of the spreading processes. 
\end{itemize}

\noindent Equation \eqref{fig: spreadingPr} is the spreading processes on networks. It states that the value of the node $i$ at time $t+1$ is the sum of its initial value and quantity that gets to node $i$ from its nearest neighbours in time interval $\delta t$; such amount is proportional to the difference of the values. 

%\medskip
%\noindent How spreading on network occurs and how long it lasts depend of some factors. %Interconnectedness of nodes in the network, whether strategic or random spreading, resistance %ability of the nodes and infectiousness likelihood are some factors that affect spreading on %network.  
\noindent The main idea of the spreading processes is at the difference between the values of a node and its neighbour \cite{zhukov2015diffusion}. We briefly discuss random walk as a model of spreading processes on networks \cite{lovasz1993random}.

\subsection{Random Walk on Networks}

\begin{definition}[Random Walk]
Let $G(N,p)$ be a random network. A random walk on a network is a sequence of nodes $v_{0}, v_{1}, v_{2}, ... v_{k}$ such that $v_{k+i}$ is chosen to be random neighbour $v_{k}$; $(v_{k}, v_{k+1})\in E(G)$ \cite{lovasz1993random}.
\end{definition}

\begin{figure}[H]
    \centering
    \begin{subfigure}[b]{0.49\textwidth}
    \centering
    \includegraphics[width=\textwidth]{Images/Undirected_network.png}
    \caption{A Connected Undirected and Unweighted Network}
    \label{fig:graphforRand}
    \end{subfigure}
    \hfill
    \centering
    \begin{subfigure}[b]{0.49\textwidth}
    \centering
    \includegraphics[width=\textwidth]{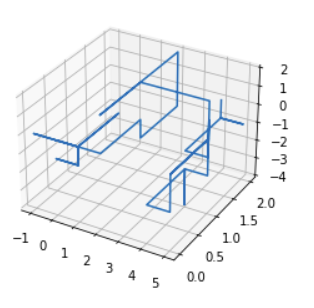}
    \caption{Random Walk on a Lattice}
    \label{fig:RandWak}
    \end{subfigure}
\end{figure} 

\noindent We consider an undirected and unweighted network in Figure \eqref{fig:graphforRand}. If a walker starts at node 1, he/she freely chooses next node of visit: node 2 or node 3. If he/she chooses to visit node 3 (without returning to node 1), then his/her probability of getting to another node is $\frac{1}{2}$. So, the random process allows him/her to visit node 4 or node 5. The random walk is the sequence of nodes visited and can be described as a stochastic movement as in Brownian motion \cite{codling2008random}. 

\medskip
\noindent By simulation of random walks on the network, the walker starts at position (0,0,0) and at each instance takes a step in either $x,y$ or $z$ direction. With 1000 walkers and 10 steps per walk, random walk -- Figure \eqref{fig:RandWak} is produced.

\medskip
\noindent The probability of the transition is given by 
$\mathbb{P}_{ij} = \mathbb{P}(v_{t+1} = v_{j}\vert v_{t} = v_{i})$,  \par
\noindent where \[\sum_{i}^{} \mathbb{P}_{ij} = 1.\]

\noindent $\mathbb{P}$ is a transition probability matrix which predicts the chance of the walker being at a node at particular time, the tendency that at next step the walker will be at node $j$ \cite{lovasz1993random}.
\noindent For the undirected and unweighted network, random walk is defined by the transition matrix 

\begin{equation}
    \mathbb{P}_{ij} =
    \begin{cases}
      \frac{1}{d(i)}, & \text{if} \ \exists \ (i,j)\in E(G), i, j \ \text{adjacent}\\ \
      0,& \text{otherwise}.
    \end{cases}
  \end{equation}
\noindent $d(i)$ is the degree of node $i$ which is possible since every node in the network G has at least one degree.

\subsection{Financial Contagion as a Spread Process}

\noindent \cite{quail2011financial} traces original use of the term "financial contagion" to field of epidemiology. It involves mechanism of transmission from one infected victim to other potential victims. Financial contagion is a distress spread in connected institutions which are dealing with currency, economy or banks. It can quickly affect business sectors and entire global market. Effects of shocks in highly connected network structure are reduced when compare to low connected network \cite{gai2010contagion} but in infectious disease spread, a more connected system causes the contagious disease to disseminate fast which is more injurious to the system \cite{quail2011financial}. 

\medskip
\noindent In financial network some nodes have more influence than others. For example, 2008 global financial crises was triggered by systematically important countries (USA is among) and all other countries in the world were affected \cite{del2020multiplex}. Cascade effects, that is, propagating failure (dynamics failure) present in financial contagion, which leads to the collapse of one financial institution, can lead to another member's collapse in the network.

\section{Contagion on Financial Networks: Implementation Details and Results}

\noindent In this section we use 100 banks (only 85\% are solvent) as nodes and 2 ranges of probability values with 15 and 20 equally spaced intervals which determine the rate of having a link between any 2 banks. We evaluate the effects of the varying probability values and percentage solvency and default in the financial network. Also, we illustrate and analyze the results of the implementation. 

\subsection{Implementation of the Erdős–Rényi Model}

\noindent We implement E–R model using 100 banks out of which 15 are shocked and the shocks propagate in time until they default. There are 2 ranges of probability values: 0.03 to 0.10 and 0.06 to 0.13 respectively. The network is directed, weighted and randomly generated. The incoming and outgoing links denote assets and liabilities respectively between banks; the directed-weighted links indicate that inter-banks' exposures consist of the assets and liabilities. The ranges of the probability values have 15 and 20 equally spaced intervals. Investigation of the correlation between the varying probability values and percentage solvency and default in the network is evaluated.

\medskip
\noindent Since random graphs are used, we generate average of ranges of 10 and 20 times random iterations of the contagions  trials for "more true results" and proper analysis. 

\section{Results and Analysis: Probability Values, Percentage Solvency and Default}

\begin{figure}[H]
    \centering
    \begin{subfigure}[b]{0.49\textwidth}
    \centering
     \includegraphics[width=\textwidth]{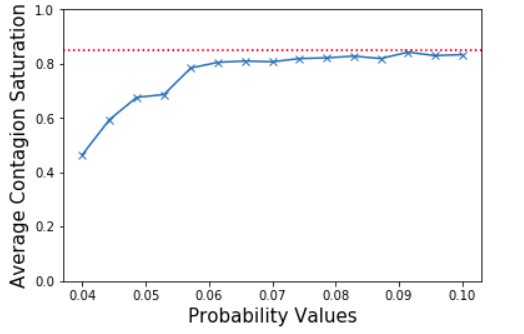}
    \caption{Probability of Interconnectedness of Banks with Average Percentage of Solvent Banks from 10 Random Iterations}
    \label{fig:probVpercents.04to.10}
    \end{subfigure}
    \hfill
    \centering
    \begin{subfigure}[b]{0.49\textwidth}
    \centering
    \includegraphics[width=\textwidth]{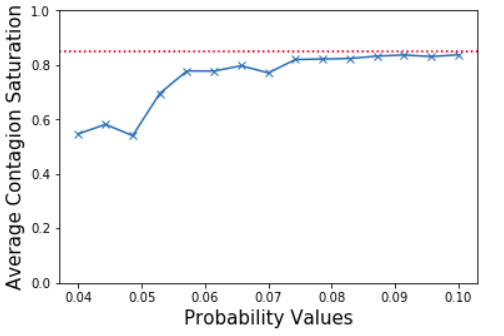}
    \caption{Probability of Interconnectedness of Banks with Average Percentage of Solvent Banks from 20 Random Iterations}
    \label{fig:20Intforprob0.04to0.10}
    \end{subfigure}
\end{figure}

\begin{figure}[H]
    \centering
    \begin{subfigure}[b]{0.49\textwidth}
    \centering
    \includegraphics[width=\textwidth]{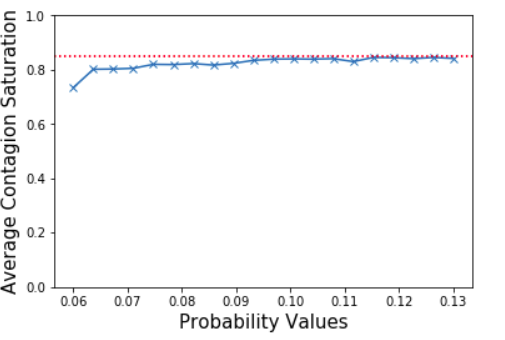}
    \caption{Probability of Interconnectedness of Banks with Average Percentage of Solvent Banks from 10 Random Iterations}
    \label{fig:probVpercen}
    \end{subfigure}
    \hfill
    \centering
    \begin{subfigure}[b]{0.49\textwidth}
    \centering
    \includegraphics[width=\textwidth]{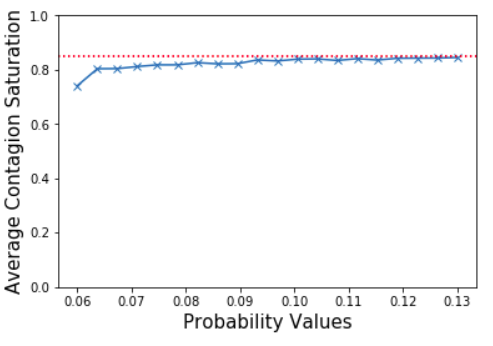}
    \caption{Probability of Interconnectedness of Banks with Average Percentage of Solvent Banks from 20 Random Iterations}
    \label{fig:20Intforprob0.06to0.13}
    \end{subfigure}
\end{figure}

\noindent Figures \eqref{fig:probVpercents.04to.10},\eqref{fig:20Intforprob0.04to0.10},\eqref{fig:probVpercen} and \eqref{fig:20Intforprob0.06to0.13} are varying ranges of the probability values versus average of ranges of 10 and 20 times random iterations of the contagions trials in each case.
 
\noindent The financial network is set at 15.00\% default from outset -- only 85.00\% of the banks are solvent, which is indicated with the (red) dotted lines. For the range of probability values from 0.04 to 0.10 with 15 equally spaced intervals, the financial network records approximately 31.50\% (that is, $(85.00-53.50)\%$) and 32.20\% (that is, $(85.00-52.80)\%$) of the banks' default at probability value of 0.04 from average of 10 and 20 random iterations of contagion trials respectively. When any bank defaults, its liabilities are redistributed to others in the network. Low solvency regions with low probability values have high level of contagion risk because financial network with low degree of interconnectedness, net assets of many banks are considerably insufﬁcient to cover losses on interbank assets emanating from cascade effects \cite{amini2016resilience}. The gap between the (red) dotted lines -- difference between the "flat-at-top" region and the network generated from fixed range of probability values (0.04 to 0.10), indicate the magnitude of the default. 

\noindent Setting a higher range of the probability values (from 0.06 to 0.13) reveals that higher probability values hinder financial contagion (cf. Figures \eqref{fig:probVpercen} and \eqref{fig:20Intforprob0.06to0.13}). The gap between the (red) dotted lines and the network generated from fixed range of probability values (0.06 to 0.13) have become closer. 2.50\% increase in the range of the probability values reduces the default to 11.55\% (that is, $(85.00-73.45)\%$) and 11.00\% (that is, $(85.00-74.00)\%$) at probability value of 0.06 from average of 10 and 20 times random iterations of contagion trials respectively. 

\medskip
\noindent The peak of solvency regions in both random iterations can be described as safe region that can check banks' collapse since it distresses contagion risk \cite{abduraimova2021solvency}. The high solvency region has low tendency of defaults. Epidemic (for example, Covid-19 pandemic) or huge macroeconomic downturn can invalidate our argument but policymakers should safeguard financial institutions during such economic upheaval. 

\medskip
\noindent The effect of probability of interconnectedness of banks in a financial network on solvency or default of the system can as well be verified by keeping probability values fixed. The number of banks at default are varied which then yields varied degrees of each bank in the network. 

\medskip
\noindent Let us examine data of average number of the solvent banks and probability values with 15 and 20 equally spaced intervals.
 
\begin{figure}[H]
    \centering
    \begin{subfigure}[b]{0.49\textwidth}
    \centering
    \includegraphics[width=\textwidth]{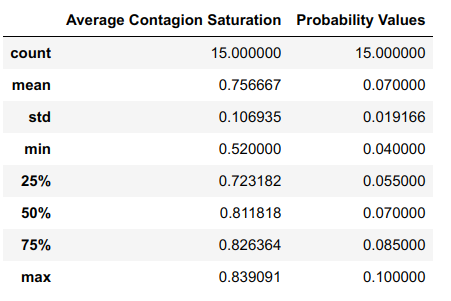}
    \caption{Summary Statistics of Probability of Interconnectedness of Banks and Average Percentage of Solvent Banks from 10 Random Iterations}
    \label{fig:probVper}
    \end{subfigure}
    \hfill
    \centering
    \begin{subfigure}[b]{0.49\textwidth}
    \centering
    \includegraphics[width=\textwidth]{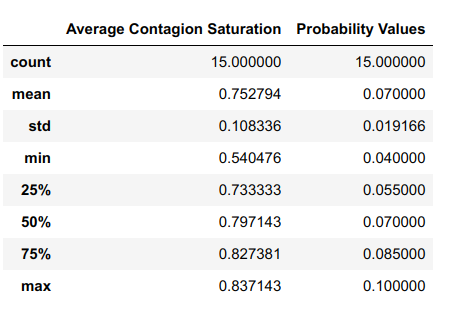}
    \caption{Summary Statistics of Probability of Interconnectedness of Banks and Average Percentage of Solvent Banks from 20 Random Iterations}
    \label{fig:summRYST0.04to0.1}
    \end{subfigure}
\end{figure} 

\begin{figure}[H]
    \centering
    \begin{subfigure}[b]{0.49\textwidth}
    \centering
    \includegraphics[width=\textwidth]{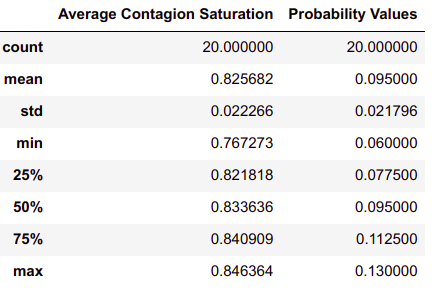}
    \caption{Summary Statistics of Probability of Interconnectedness of Banks and Average Percentage of Solvent Banks from 10 Random Iterations}
    \label{fig:summstats}
    \end{subfigure}
    \hfill
    \centering
    \begin{subfigure}[b]{0.49\textwidth}
    \centering
    \includegraphics[width=\textwidth]{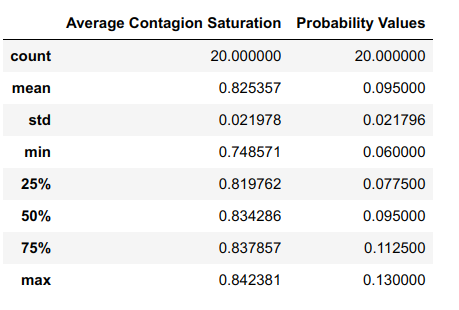}
    \caption{Summary Statistics of Probability of Interconnectedness of Banks and Average Percentage of Solvent Banks from 20 Random Iterations}
    \label{fig:summstat2}
    \end{subfigure}
\end{figure} 
\noindent From the summary statistics of mean percentage of solvent banks and different ranges of the probability values (cf. Figures \eqref{fig:probVper}, \eqref{fig:summRYST0.04to0.1}, \eqref{fig:summstats} and \eqref{fig:summstat2}), there are no statistical differences in the means of 15 and 20 equally spaced intervals of the probability values. Also, both average of ranges of 10 and 20 times random iterations of the contagions trials show no statistically significant difference. 

\medskip
\noindent Hence, we deduce that the number of grid points -- fineness of the intervals of the probability values, has no effect on evaluation of percentage solvency or default of the banks. The number of sampling of the contagion trials does not affect the average contagion saturation. The mutual sampling need not be fine grid.

\section{Conclusion}

\noindent This mini-project implements contagion in any financial network where shocks can spread. Effects of such shocks are reduced if the banks are highly connected since all banks in the network share a little bit of the shocks; the shocks' impacts become minimal. The percentage solvency and default depend on the probability values. The range of probability values and banks' percentage solvency have positive correlation. The number of grid points of the probability values and sampling of the contagion trials have no effect on determining the percentage solvency and default. 

\medskip
\noindent Epidemic or any economic upheaval can invalidate our argument. In such situation, policy makers should safeguard the financial institutions.

\medskip
\noindent Since number of banks are discrete, we use the E-R model which is approximated by Poisson distribution -- a discrete probability distribution. A good direction in future work can be done by using stochastic method of modelling real-world network.  The model is stochastic block model or E-R mixture model.

\bibliographystyle{apacite}%\bibliographystyle{unsrtnat}
\bibliography{references}  %%% Uncomment this line and comment out the ``thebibliography'' section below to use the external .bib file (using bibtex) .

%\begin{thebibliography}{1}




% \end{thebibliography}

\vskip 0.4in

\noindent\textbf{Acknowledgements}
    
\noindent Immense thanks to my mentors, \textbf{Shazia'Ayn Babul} and \textbf{Sofia Medina} of Mathematical Institute, University of Oxford, United Kingdom, during the Mfano Africa-Oxford Mathematics 2023 Research Mentorship. 
	 
\end{document}